\documentclass[preprint,12pt,num,compress,sort]{elsarticle}
\usepackage{amsmath} 
\usepackage{amssymb}
\usepackage{graphicx}
\usepackage[colorlinks=true, pdfstartview=FitV, linkcolor=blue, citecolor=red, urlcolor=blue]{hyperref}
\usepackage[compat=1.1.0]{tikz-feynhand}

\newcommand{\rmd}{\mathrm{d}}
\newcommand{\rme}{\mathrm{e}}
\newcommand{\rmi}{\mathrm{i}}
\newcommand{\bx}{\boldsymbol{x}}

\newcommand{\bnabla}{\boldsymbol{\nabla}}
\newcommand{\bOmega}{\boldsymbol{\Omega}}
\newcommand{\calD}{\mathcal{D}}

\journal{Physics Letters B}

\begin{document}

\begin{frontmatter}

\title{  \vspace*{-2em} {\small \hfill RIKEN-iTHEMS-Report-25} \\
  \vspace{1.5em}
  Perturbation theory of rotating scalar fields\\
  and vacuum insensitivity to rotation
  }

\author[TUS]{Ryo~Kuboniwa}
\author[TUS,RIKEN]{Kazuya~Mameda}

\address[TUS]{Department of Physics, Tokyo University of Science, Shinjuku, Tokyo 162-8601, Japan}
\address[RIKEN]{RIKEN iTHEMS, RIKEN, Wako, Saitama 351-0198, Japan}

\begin{abstract}
We formulate the finite-temperature perturbation theory of interacting scalar fields under external rotation.
Because of the translational non-invariance in the radial direction, Green's functions are described using the Fourier-Bessel basis, instead of the conventional Fourier basis.
We derive the leading-order correction to the partition function and the one-loop self-energy.
The Feynman rules obtained in our perturbation theory shows that due to the finite-size effect required by the causality constraint, the zero-temperature thermodynamics in the perturbation theory is unaffected by rotation, similarly to that in the noninteracting theory.
\end{abstract}


\end{frontmatter}

\section{Introduction}

Rotation is one of the external sources that yields intriguing phenomena in quantum theory.
In high-energy nuclear physics, the experimental measurement of the strongest vorticity in nature, that of quark-gluon plasma~\cite{STAR:2017ckg}, motivates the development of relativistic thermal field theory under external rotation.
For slow rotation, this framework is applied to chiral transport phenomena induced by vorticity~\cite{Vilenkin:1978hb,Vilenkin:1979ui,Amado:2011zx,Landsteiner:2011cp, Hattori:2016njk,Fukushima:2024tkz} and the spin dynamics of quantum chromodynamics (QCD)~\cite{Florkowski:2018fap,Gao:2020vbh,Becattini:2020ngo,Becattini:2024uha,Hidaka:2022dmn}.
A similar direction is also expanding into investigations in relativistic thermodynamics related to chiral symmetry breaking~\cite{Chen:2015hfc, Jiang:2016wvv, Chernodub:2016kxh,Chen:2023cjt} and the deconfinement transition~\cite{Chernodub:2020qah,Fujimoto:2021xix,Chen:2022smf}, including lattice simulation studies~\cite{Chernodub:2022veq, Braguta:2021jgn, Braguta:2023iyx,Yang:2023vsw} (see also Refs.~\cite{Huang:2017pqe,Liu:2017spl,Cao:2019ctl,Chen:2019tcp,Zhang:2018ome} for other possible phase transitions),
which focus on rigid rotation as the fundamental basis for understanding the dynamics of rotating QCD matter.

An essential feature of rotating systems is that the translation along the radial direction is broken unless angular velocity is much smaller than other scales.
This is because rotation specifies the center axis of systems, and the causality constraint requires introducing a boundary on the edge of system.
The radial inhomogeneity brings difficulties in the field-theoretical treatment of rotating matter.
In lattice gauge theory, for example, rapid rotation leads to a problem on renormalization~\cite{Yamamoto:2021oys}, in addition to the sign problem~\cite{Yamamoto:2013zwa}.
Besides, the Green's function under rotation is not easily tamable because of the necessity to employ the Fourier-Bessel basis instead of the usual Fourier basis in the Cartesian coordinate.
Indeed, apart from lattice simulations, most preceding studies are restricted to noninteracting theories or at most mean-field theories, which evades loop-diagram computations; see Ref.~\cite{Salvio:2025rma,Kawaguchi:2025mkh} for exceptions.
Although the slow-rotation limit allows us to employ the usual Green's function described with the Fourier basis, then the thermodynamics does not fully incorporate the rotational effect involving the orbital-rotation coupling.

Another important characteristic of rotating systems is the rotational effect on vacuum.
In a noninteracting theory, the finite-size effect required to respect the causality constraint prohibits any visible rotational effect at zero temperature~\cite{Vilenkin:1980zv, Davies:1996ks, Ambrus:2015lfr}.
While this is analogous to the Silver-Blaze phenomenon, where the mass threshold overcomes the real chemical potential, its underlying mechanism is a more subtle competition between two mode-dependent quantities: the infrared energy gap $\epsilon_\mathrm{IR}$ and the rotation-induced effective chemical potential $\mu_\mathrm{rot}$~\cite{Ebihara:2016fwa}.
A natural but nontrivial question is, hence, whether this vacuum insensitivity to rotation is inherited by an interacting theory;
interactions allowing angular-momentum exchange at vertices could in principle alter the subtle competition between $\epsilon_\mathrm{IR}$ and $\mu_\mathrm{rot}$.
Revealing the correct rotational response of the vacuum in an interacting theory is essential for field-theoretical studies of rotating matter, including the impact of the finite-size effect in rotating systems.

In this paper, we formulate the finite-temperature perturbation theory for the $\lambda\phi^4$ theory under rigid rotation by employing the Fourier-Bessel basis.
This perturbation theory shows that due to the finite-size effect, the vacuum insensitivity to rotation is observed up to $\mathcal{O}(\lambda^2)$.
Our formulation for the spinless field not only becomes a foundation on analyses based on any other theories and models, but also clarifies the contribution from the orbital-rotation coupling.

\section{Noninteracting theory}

One of the approaches to introduce the rotational effect on thermodynamics is to follow the maximum entropy principle~\cite{Landau:1980mil}.
In systems with rotational symmetry, the thermal ensemble acquires angular velocity as the Lagrange multiplier corresponding to the angular momentum conservation~\cite{Vilenkin:1980zv}.
Therefore, the partition is written as
\begin{equation}
\label{eq:partition-function}
 Z = \mathrm{tr} \Bigl[\rme^{-\beta (H-\Omega L)}\Bigr],
\end{equation}
where $\beta = T^{-1}$ is the inverse temperature and $\bOmega = \Omega \hat{z}$ is the angular velocity;
in this paper, we employ a constant $\Omega\geq 0$.
In addition, $H$ and $L$ are the field operators of Hamiltonian and angular momentum parallel to $\bOmega$.
The above expression indicates the rotational effect plays a similar role to the finite-density effect.

Let us first focus on the noninteracting real scalar theory, which is described by the following Lagrangian:
\begin{equation}
\begin{split}
 \mathcal{L}_0
 = \frac{1}{2} \partial_\mu \phi \partial^\mu \phi
  - \frac{1}{2} m^2 \phi^2
\end{split}
\end{equation}
with $m$ being the mass.
The Hamiltonian density is obtained from the Legendre transformation as $\mathcal{H} = \pi \partial_t \phi - \mathcal{L}$ with $\pi = \partial\mathcal{L}/\partial(\partial_t \phi)$, and thus the integrated one reads
\begin{equation}
\begin{split}
 H_0 
 &= \int\rmd^3 \bx\,\mathcal{H}_0 \\
 \quad
 &= \int \rmd^3 \bx
 \biggl[ \frac{1}{2} \pi^2 + \frac{1}{2}(\bnabla\phi)^2 + \frac{1}{2}m^2\phi^2 \biggr].
\end{split}
\end{equation}
The canonical angular momentum is given as
\begin{equation}
\begin{split}
 L
 &=\int \rmd^3 \bx (x T^{02} - y T^{01}) \\
 &= -\int \rmd^3 \bx \,\pi(x\partial_y -y\partial_x)\phi
\end{split}
\end{equation}
with the canonical energy-momentum tensor defined as the Noether current for the translational symmetry: $T^{\mu\nu} = [\partial\mathcal{L}/\partial(\partial_\mu \phi)]\partial^\nu\phi - \eta^{\mu\nu}\mathcal{L}$, where $\eta_{\mu\nu} = \mathrm{diag}(1,-1,-1,-1)$.
Inserting an infinite number of the complete sets $\int \rmd \phi |\phi\rangle \langle\phi| =1 $ and $
\int \rmd \pi/(2\pi) |\pi\rangle \langle\pi| = 1$, we rewrite the partition function of the noninteracting theory into the following path-integral form:
\begin{equation} 
\begin{split}
\label{eq:Z0}
 Z_0 
 &= \mathrm{tr} \Bigl[\rme^{-\beta (H_0-\Omega L)}\Bigr]\\
 &= \int [\rmd \pi] \int [\rmd \phi] \exp \biggl[ -\int_x
 \biggl(
 \frac{1}{2} \pi^2 - \rmi\pi \partial_\tau \phi \\
 &\quad + \Omega \pi (x\partial_y -y\partial_x)\phi
 + \frac{1}{2}(\bnabla \phi)^2 + \frac{1}{2} m^2 \phi^2
 \biggr)\biggr] \\
 &= \int [\rmd \phi]\, \rme^{S_0}
\end{split}
\end{equation}
with $\int_x = \int_0^\beta \rmd\tau \int \rmd^3 \bx$.
In the cylindrical coordinate $x^\mu = (r,\theta,z,\tau)$, the action $S_0$ reads
\begin{equation}
\begin{split}
\label{eq:action0}
 S_0 
 &=-\frac{1}{2} \int_x \phi \biggl[ -(\partial_\tau + \rmi \Omega\partial_\theta)^2 \\
 &\qquad
 -\partial_r^2 - \frac{1}{r} \partial_r - \frac{1}{r^2} \partial_\theta^2 - \partial_z^2
  + m^2\biggr]\phi.
\end{split}
\end{equation}

To perform the above functional integration, we diagonalize the derivative operator in Eq.~\eqref{eq:action0}.
The eigenfunction is represented with the same form as that for $\Omega=0$ because of the rotational symmetry.
An important remark is that we here need to carefully take into account the boundary condition along the radial direction so that the causality constraint is respected: $\Omega R\leq 1$ with $R$ being the radius of the cylindrical system.
When we adopt the Dirichlet boundary condition at the edge of boundary, our basis function to diagonalize the derivative operator is expressed as
\begin{equation}
\begin{split}
\label{eq:varphi}
 &\varphi_{\nu}(x) = \sqrt{\frac{\beta}{V}} \, \rme^{\rmi \omega_n \tau + \rmi q z + \rmi l \theta} \frac{J_l(p_{l,k} r)}{N_{l, k}},\\
 &\quad p_{l,k} = \frac{\xi_{l,k}}{R},
 \quad
 N_{l, k} = |J_{l+1}(\xi_{l,k})|
\end{split}
\end{equation}
with $\xi_{l,k} = \xi_{-l,k}$ being the $k$-th positive root of the Bessel function $J_l(\xi)$,  the system volume $V=\pi R^2 h$, and the longitudinal system size $h$.
The basis function $\varphi_\nu$ is designated by the set of quantum numbers as
\begin{equation}
\label{eq:nu}
\nu =  \{\nu_\parallel,\nu_\perp\},\quad \nu_\parallel= \{n, q\},
\quad \nu_\perp =\{l,k\}
\end{equation}
with the bosonic Matsubara frequency $\omega_n = 2 n\pi T$, the continuous longitudinal momentum $q$, and the $z$-component of angular momentum $l$.
The normalization in Eq.~\eqref{eq:varphi} follows from the orthogonal relation
\begin{equation}
\label{eq:orthogonal}
\int_x \varphi^*_{\nu'} (x) \varphi_\nu (x) 
= \beta^2 \delta_{\nu,\nu'},
\end{equation}
where
\begin{equation}
\delta_{\nu,\nu'}
=\delta_{\nu_\parallel,\nu_\parallel'}\delta_{l,l'}\delta_{k,k'},\quad
\delta_{\nu_\parallel,\nu_\parallel'} =\delta_{n,n'}\cdot \frac{2\pi}{h}\delta(q-q').
\end{equation}
One can readily show Eq.~\eqref{eq:orthogonal} with the spatial integral over the transverse plane
\begin{equation}
 \int \rmd^2 \boldsymbol{r}\, \rme^{\rmi(l-l')\theta}  J_l(p_{l, k}r) J_{l'}(p_{l', k'}r)
 = \pi R^2 N_{l, k}^2 \, \delta_{l, l'}\delta_{k, k'}.
\end{equation}

Let us expand the real scalar field in terms of the above basis:
\begin{equation}
\label{eq:FB-expansion}
 \phi(x)
 = \sum_{\nu} \tilde{\phi}_{\nu} \, \varphi_{\nu}(x),
\end{equation}
where we denote the sum as
\begin{equation}
 \sum_\nu = \sum_{n=-\infty}^{\infty}\sum_{l=-\infty}^\infty\sum_{k=1}^\infty\, h\int_{-\infty}^{\infty}\frac{\rmd q}{2\pi} .
\end{equation}
Using the orthogonal relation~\eqref{eq:orthogonal}, we diagonalize the action $S_0$ as
\begin{equation}
\begin{split}
 S_0
 &= -\frac{\beta^2}{2}\sum_\nu 
\Bigl[(\omega_n-\rmi\Omega l)^2 + \epsilon_\nu^2 \Bigr]|\tilde{\phi}_{\nu}|^2
\end{split}
\end{equation}
with the energy-dispersion relation
\begin{equation}
 \epsilon_\nu = \sqrt{m^2 + q^2 + p_{l, k}^2} .
\end{equation}
After carrying out the Gaussian integration in terms of $|\tilde{\phi}_\nu|$ in Eq.~\eqref{eq:Z0} and then the Matsubara summation, we arrive at
\begin{equation}
\begin{split}
\label{eq:lnZ0}
 \ln Z_0 
 &= -\frac{V}{2\pi R^2} \sum_{l, k} \int \frac{\rmd q}{2\pi} \\
 &\quad\times \left[\beta \epsilon_\nu + 2\ln \left\{1 - \rme^{-\beta (\epsilon_\nu - \Omega l)} \right\} \right] .
\end{split}
\end{equation}
Here, the factor $2$ in the thermal part is from the degenerate contributions of particle and antiparticle, which are not distinguished by rotation.
Also, the above $\ln Z_0$ is invariant for $\Omega \to -\Omega$ as the sum of $l$ runs from $-\infty$ to $\infty$, indicating that the partition function itself is irrelevant to the direction of rotation.

We argue the importance of the boundary condition and the causality constraint $\Omega R \leq 1$.
From the inequality $\xi_{l, 1} > |l|$ for arbitrary $l$~\cite{GIORDANO1983221}, we show~\cite{Vilenkin:1980zv}
\begin{equation}
\begin{split}
\label{eq:positivity}
 \epsilon_\nu - \Omega |l|
 &\ge \frac{\xi_{l, k}}{R} - \Omega|l| \\
 &\ge \frac{1}{R}(\xi_{l, k} - |l|) 
 > 0.
\end{split}
\end{equation}
The inequality above ensures that the partition function in Eq.~\eqref{eq:lnZ0} always takes a real value, and thus the thermodynamics of noninteracting rotating scalars is well-defined.
At the same time, the $\Omega$-dependence disappears in $\ln Z_0$ at zero temperature, as $\ln\bigl[1 - \rme^{-\beta (\epsilon_\nu - \Omega l)} \bigr]\to 0$.
This property is shared with arbitrary thermodynamic quantities described by the Bose distribution
\begin{equation}
 n_\mathrm{B}(\nu) = \frac{1}{\rme^{\beta (\epsilon_\nu - \Omega l)}-1},
 \quad
 \bar{n}_\mathrm{B}(\nu) = \frac{1}{\rme^{\beta (\epsilon_\nu + \Omega l)}-1},
\end{equation}
which vanish in the zero-temperature limit.
Therefore, the rotational effect is visible at finite temperature, but invisible at zero temperature.
The physical interpretation is that while thermal excited modes can be affected by external rotation, the vacuum, which has nothing to rotate, cannot.
The same property also appears in fermionic systems~\cite{Ambrus:2015lfr,Ebihara:2016fwa}.

\section{Correction to partition function}
We now analyze the interacting scalar theory, which is described by the Lagrangian
\begin{equation}
 \mathcal{L} = \mathcal{L}_0 - \lambda\phi^4 .
\end{equation}
Similarly to the noninteracting case, the corresponding partition $Z$ is expressed by the functional integral over $\phi$.
From the expansion in terms of $\lambda$, the first order correction to $\ln Z$ reads
\begin{equation}
\begin{split}
 \label{eq:lnZ1-1}
 \ln Z_1 &= \frac{-\lambda \int_x \int [\rmd \phi] \rme^{S_0} \phi^4 (x)}{\int [\rmd \phi] \rme^{S_0}} \\
 &= -\lambda \int_x \sum_{\nu_1, \nu_2, \nu_3, \nu_4} \varphi_{\nu_1} \varphi_{\nu_2} \varphi_{\nu_3} \varphi_{\nu_4} \, \frac{A}{B} .
\end{split}
\end{equation}
In the second line, we expanded $\phi$ as Eq.~\eqref{eq:FB-expansion} with 
\begin{equation}
\begin{split}
 A &= \biggl[\prod_{\nu}
 \int \rmd \tilde{\phi}_{\nu}  G_{\nu}\biggr] \tilde{\phi}_{\nu_1} \tilde{\phi}_{\nu_2} \tilde{\phi}_{\nu_3} \tilde{\phi}_{\nu_4}, \\
 B & = \prod_{\nu} \int \rmd \tilde{\phi}_{\nu} G_{\nu},
\end{split}
\end{equation}
where $G_\nu$ is the Gaussian function
\begin{equation}
\begin{split}
 G_\nu
 = \exp \left \{ -\frac{\beta^2}{2} [(\omega_{n} -\rmi\Omega l)^2 + \epsilon_\nu^2] \, |\tilde{\phi}_{\nu}|^2 \right \}.
\end{split}
\end{equation}
The numerator $A$ vanishes unless $\nu_{3\parallel}= -\nu_{1\parallel}$, $l_3 = -l_1$, $k_3 = k_1$,  $\nu_{4\parallel}= -\nu_{2\parallel}$, $l_4 = -l_2$, and $k_4 = k_2$, or the other two permutations thereof.
Then, the summation over $\nu_3$ and $\nu_4$ yields the symmetrization factor $3$, as follows:
\begin{equation}
 \sum_{\nu_3,\nu_4} \varphi_{\nu_3} \varphi_{\nu_4} A
 = 3 \biggl[\prod_\nu \int \rmd\tilde{\phi}_\nu G_\nu\biggr]\varphi_{\nu_1}^* \varphi_{\nu_2}^* |\tilde\phi_{\nu_1}|^2 |\tilde\phi_{\nu_2}|^2 .
\end{equation}
Eventually, Eq.~\eqref{eq:lnZ1-1} is evaluated as
\begin{equation}
\begin{split}
\label{eq:lnZ1-2}
 \ln Z_1
 &= -\frac{3\lambda T}{V} \sum_{\nu_1, \nu_2} I_{\nu_{1\perp},\nu_{1\perp},\nu_{2\perp},\nu_{2\perp}} \, \mathcal D_0(\nu_1) \, \mathcal D_0(\nu_2),
\end{split}
\end{equation}
where we define the propagator in frequency momentum space as
\begin{equation}
\label{eq:calD_0}
 \calD_0 (\nu)
 = \frac{1}{(\omega_n -\rmi\Omega l)^2 + \epsilon_\nu^2} .
\end{equation}
The stark difference from the expression in the Cartesian coordinate is found in the radial integral
\begin{equation}
\begin{split}
\label{eq:I}
 I_{\nu_{1\perp},\nu_{2\perp},\nu_{3\perp},\nu_{4\perp}}
 &= \frac{2}{R^2} \int_0^R \rmd r \, r\prod_{i=1}^4
 \frac{J_{l_i}(p_{l_i,k_i} r)}{N_{l_i, k_i}} .
\end{split}
\end{equation}
This integral involving the product of four $J_{l}(p_{l, k} r)$ is due to the four-point interaction vertex.
To our best knowledge, no analytical formula is known for this integral, even in the $R\to\infty$ limit~\footnote{%
For $R\to\infty$, Ref.~\cite{cosmin2009} argues the Mellin transformation to the integral containing the four-product of Bessel functions, extending to the three-product case~\cite{tyler1990analysis}.
We have, however, numerically confirmed that the formula does not provide correct values at least for the integral~\eqref{eq:I}.%
}.
Hence, all of the computation involving the radial integral~\eqref{eq:I} is performed numerically.
Also, we note that if the eigenfunction were described by a plane wave $\exp(\rmi p_x x+\rmi p_y y)$ instead of Eq.~\eqref{eq:varphi}, the perturbative correction to the partition function~\eqref{eq:lnZ1-2} could be reduced to a much simpler form since the phase space sum over $\nu_1$ and $\nu_2$ are totally decoupled as
\begin{equation}
\label{eq:Cartesian}
\begin{split}
  &\frac{1}{(\pi R^2)^2} \sum_{\nu_{1\perp},\nu_{2\perp}} I_{\nu_{1\perp},\nu_{1\perp},\nu_{2\perp},\nu_{2\perp}} \\
  &\quad \to \int \frac{\rmd^2 \boldsymbol{p}_{1\perp}}{(2\pi)^2}\int \frac{\rmd^2 \boldsymbol{p}_{2\perp}}{(2\pi)^2},
\end{split}
\end{equation}
where $\boldsymbol{p}_{i\perp}$ is the radial momentum in the continuum limit.

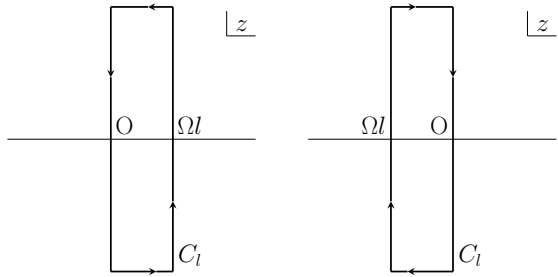
\begin{figure}
 \begin{minipage}{0.45\columnwidth}
     \centering
    \scalebox{0.5}{
     \begin{tikzpicture}
       \draw[semithick] (-2.5, 0) -- (3.5, 0) node[right]{};
       \draw (0, 0) node[above right, font=\LARGE]{O};
       \draw (1.5, 0) node[above right, font=\LARGE]{$\Omega l$};
       \draw[semithick] (2.8, 2.5) -- (3.5, 2.5);
       \draw[semithick] (2.8, 2.5) -- (2.8, 3.1);
       \draw (2.8, 2.5) node[above right, font=\LARGE]{$\,z$};
  \draw[->,>=stealth, very thick] (1.5, -3.2) -- (1.5, -1.5);
  \draw[very thick] (1.5, -1.5) -- (1.5, 3.2);
  \draw[->,>=stealth, very thick] (1.5, 3.2) -- (0.9, 3.2);
  \draw[very thick] (0.9, 3.2) -- (0, 3.2);
  \draw[->,>=stealth, very thick] (0, 3.2) -- (0, 1.5);
  \draw[very thick] (0, 1.5) -- (0, -3.2);
  \draw[->,>=stealth, very thick] (0, -3.2) -- (1.1, -3.2);
  \draw[very thick] (1.1, -3.2) -- (1.5, -3.2) node [above right, font=\LARGE]{$C_l$};
 \end{tikzpicture}
 }
 \end{minipage}
 \begin{minipage}{0.45\columnwidth}
     \centering
    \scalebox{0.5}{
     \begin{tikzpicture}
       \draw[semithick] (-3.5, 0) -- (2.5, 0) node[right]{};
       \draw (0, 0) node[above left, font=\LARGE]{O};
       \draw (-1.5, 0) node[above left, font=\LARGE]{$\Omega l$};
       \draw[semithick] (1.8, 2.5) -- (2.5, 2.5);
       \draw[semithick] (1.8, 2.5) -- (1.8, 3.1);
       \draw (1.8, 2.5) node[above right, font=\LARGE]{$\,z$};
  \draw[->,>=stealth, very thick] (-1.5, -3.2) -- (-1.5, -1.5);
  \draw[very thick] (-1.5, -1.5) -- (-1.5, 3.2);
  \draw[->,>=stealth, very thick] (-1.5, 3.2) -- (-0.9, 3.2);
  \draw[very thick] (-0.9, 3.2) -- (0, 3.2) ;
  \draw[->,>=stealth, very thick] (0, 3.2) -- (0, 1.5);
  \draw[very thick] (0, 1.5) -- (0, -3.2) node [above right, font=\LARGE]{$C_l$};
  \draw[->,>=stealth, very thick] (0, -3.2) -- (-1.1, -3.2);
  \draw[very thick] (-1.1, -3.2) -- (-1.5, -3.2) ;
 \end{tikzpicture}
 }
 \end{minipage}
 \caption{Contour $C_l$ for $l> 0$ (left) and $l<0$ (right).}
\label{fig:contour}
\end{figure}

To reduce Eq.~\eqref{eq:lnZ1-2}, we compute
$T\sum_n \calD_0(\nu)$ in the parallel manner to the usual Cartesian case;
the radial integral $I$ is irrelevant to the sum.
Converting the sum to a contour integral with the replacement $\omega_n-\rmi\Omega l\to -\rmi z$, we rewrite the Matsubara sum into
\begin{equation}
\begin{split}
 \label{eq:Matsubara}
 &T \sum_{n} \calD_0(\nu) \\
 &= \frac{n_\mathrm{B}(\nu)+\bar{n}_\mathrm{B}(\nu)}{2 \epsilon_\nu}
 + \biggl(\int_{-\rmi\infty}^{\rmi\infty} +\oint_{C_l}\biggr) \frac{\rmd z}{2 \pi \rmi}\frac{1}{\epsilon_\nu^2 - z^2},
\end{split}
\end{equation}
where $C_l$ is the $l$-dependent contour shown in Fig.\allowbreak~\ref{fig:contour}.
The first term on the right-hand side represents the thermal contribution, and the integral along the imaginary axis is the $\Omega$-independent and zero-temperature part.
In the usual case, this contribution is compensated by adding the counter term $-\frac{1}{2}\delta m^2\phi^2$, and renormalized into the mass%
~\footnote{%
As we see later, the leading order self-energy~\eqref{eq:Pi_r} is dependent on external momentum unlike the Cartesian case~\eqref{eq:Pi_c}.
This modification is not an obstacle to performing renormalization.
}.
We here just follow such a standard scheme, and hereafter neglect the contribution at $T=0$ and $\Omega=0$.

The most important finding in Eq.~\eqref{eq:Matsubara} is about the last integral along $C_l$.
This is independent of $T$ but dependent on $\Omega$, and thus is responsible for the rotational effect on vacuum.
However, this integral always vanishes because the poles $z=\pm\epsilon_\nu$ are located outside the contour $C_l$, due to Eq.~\eqref{eq:positivity}.
Hence, we find that at the leading order of the perturbation theory, the thermodynamics involves no visible rotational effect at zero temperature, as is in the noninteracting theory.

Eventually, the correction to the partition function is
\begin{equation}
\begin{split}
\label{eq:lnZ-3}
    \ln Z_1
    &= -3\lambda \beta V\Biggl[\prod_{i=1,2} 
    \frac{1}{\pi R^2} \sum_{\nu_{i\perp}} \int \frac{\rmd q_i}{2\pi}
    \frac{n_\mathrm{B}(\nu_i)}{\epsilon_{\nu_i}}\Biggr]\\
    &\quad\times 
    I_{\nu_{1\perp},\nu_{1\perp},\nu_{2\perp},\nu_{2\perp}}.
\end{split}
\end{equation}
In the Cartesian case, this correction is, through the replacement~\eqref{eq:Cartesian}, reduced to $\ln Z_1 \to -\frac{\lambda}{48} V T^3$ after performing the three-dimensional momentum integration~\cite{Kapusta:2006pm}.

Here, we briefly comment on the infrared divergence.
In the Cartesian perturbation theory for $m=0$, the zero mode of the Matsubara frequency brings the infrared divergence in the free propagator.
This requires the resummation of the ring-diagram series~\cite{Gell-Mann:1957wvx}, leading to the $\mathcal{O}(\lambda^{3/2})$ contributions to the self-energy and the partition function~\cite{Kapusta:2006pm}.
However, in the cylindrical coordinate with finite radius $R<\infty$, the infrared energy gap $p_{l,1}=\xi_{l,1}/R >0$ is already installed.
Therefore, the resummation is unnecessary even for $m=0$, and the next leading order is $\mathcal{O}(\lambda^2)$.

\section{Self-energy}
Let us compute the self-energy of the rotating scalars.
In the coordinate-space, the full propagator is defined as 
\begin{equation}
\label{eq:Dx}
 D(x_1,x_2)
 = \frac{\int [\rmd\phi] \phi(x_1)\phi(x_2)\, \rme^{S}}{\int[\rmd\phi]\,\rme^{S}} .
\end{equation}
The lowest-order term for $\lambda=0$ reads
\begin{equation}
 D_0(x_1,x_2)= \frac{1}{\beta^{2}}\sum_\nu \calD_0(\nu) \varphi_\nu^*(x_1)\varphi_\nu(x_2)
\end{equation}
with the momentum representation of the free propagator, $\calD_0(\nu)$ in Eq.~\eqref{eq:calD_0}, as it should.
The expansion of both the numerator and the denominator in terms of $\lambda$ provides the perturbative corrections.
Up to the leading-order, we get
\begin{equation}
\begin{split}
&D(x_1, x_2)\\
&\simeq D_0(x_1, x_2)
- 12\lambda \int_x D_0(x_1, x) D_0(x, x) D_0(x, x_2) .
\end{split}
\end{equation}
Applying the Fourier-Bessel expansion~\eqref{eq:FB-expansion} with respect to both $x_1$ and $x_2$, we obtain the frequency-momentum representation
\begin{equation}
\begin{split}
\label{eq:calD}
\calD (\nu_1,\nu_2)
&=\frac{1}{\beta^2}\int_{x_1}\int_{x_2}  \varphi_{\nu_1}(x_1) \varphi^*_{\nu_2}(x_2) D(x_1,x_2) \\
&\simeq \mathcal{D}_0(\nu_1) \, \delta_{\nu_1, \nu_2} - \mathcal{D}_0(\nu_1) \Pi_1(\nu_1, \nu_2) \mathcal{D}_0(\nu_2),
\end{split}
\end{equation}
where
\begin{equation}
\begin{split}
\label{eq:Pi_offdiag}
&\Pi_1(\nu_1, \nu_2)\\
&= \frac{12\lambda T}{V} \delta_{\nu_{1\parallel}, \nu_{2\parallel}}\delta_{l_1,l_2} \sum_\nu \mathcal{D}_0(\nu) \, I_{\nu_{1\perp},\nu_{2\perp},\nu_{\perp},\nu_{\perp}}.
\end{split}
\end{equation}
In the same manner as before, the Matsubara sum~\eqref{eq:Matsubara} reduces the function $\Pi_1$ to
\begin{equation}
\begin{split}
\label{eq:Pi_r}
\Pi_1(\nu_1, \nu_2) &= \delta_{\nu_{1\parallel}, \nu_{2\parallel}}\delta_{l_1,l_2} \Pi_{1\text{r}}(\nu_{1\perp},\nu_{2\perp}),\\
 \Pi_{1\text{r}}(\nu_{1\perp},\nu_{2\perp})
 &= \frac{12\lambda}{\pi R^2} \sum_{\nu_\perp} \int \frac{\rmd q}{2\pi}\frac{n_\mathrm{B}(\nu)}{\epsilon_{\nu}} I_{\nu_{1\perp},\nu_{2\perp},\nu_{\perp},\nu_{\perp}},
\end{split}
\end{equation}
which involves no rotational effect at zero temperature.

This $\Pi_1$ is regarded as the self-energy under rotation, but with two different structures from the Cartesian one.
One difference is that $\Pi_{1}$ depends on the external momenta $l_1=l_2,k_1,k_2$ (note that $l_1\neq l_2$ is excluded by $\delta_{l_1,l_2}$), while the Cartesian one-loop self-energy takes the constant form as
\begin{equation}
\label{eq:Pi_c}
 \Pi_{1\text{c}} = \lambda T^2.
\end{equation}
Another is about the matrix structure.
Although the conservations of frequency, momentum along $z$, and angular momentum are reflected by $\delta_{\nu_{1\parallel}, \nu_{2\parallel}}$ and $\delta_{l_1,l_2}$, respectively, the counterpart of the radial momentum is not involved here due to translational non-invariance.
That is, the self-energy has off-diagonal elements in terms of $k_1$ and $k_2$, making the propagator~\eqref{eq:calD} off-diagonal as well, while the free propagator~\eqref{eq:calD_0} is diagonal.
This off-diagonal property also modifies several computational schemes used in the Cartesian coordinate.
For instance, while the diagonal element $\Pi_{1\mathrm{r}}(\nu_\perp,\nu_\perp)$ can be derived from the functional derivative $\delta\ln Z_1/\delta\calD_0(\nu)$~\cite{Kapusta:2006pm}, the off-diagonal element cannot.

\begin{figure}
\begin{minipage}{0.5\columnwidth}
    \includegraphics[width=0.9\columnwidth]{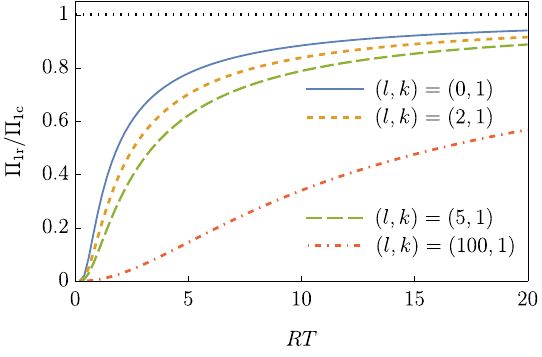}
\end{minipage}
\begin{minipage}{0.5\columnwidth}
    \includegraphics[width=0.9\columnwidth]{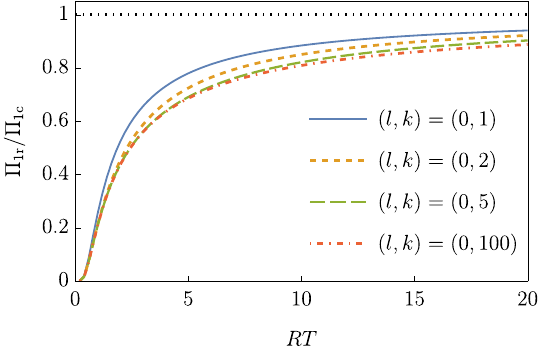}
\end{minipage}   
 \caption{Diagonal elements of the self-energy for $\Omega = 0$.}
\label{fig:Pi_d}
\end{figure}

In Fig.~\ref{fig:Pi_d}, we make the plots of the diagonal self-energy $\Pi_{1\mathrm{r}}(l,k,l,k)$ for $\Omega = m = 0$.
The vertical axis is normalized by the Cartesian self-energy~\eqref{eq:Pi_c}, and the horizontal axis is $RT$, which is the unique dimensionless parameter here.
The left (right) panel is the plot with $k=1$ fixed ($l=0$ fixed).
We observe a monotonically increasing behavior as $R$ increases and the convergent tendency $\Pi_{1\mathrm{r}}\to \Pi_{1\mathrm{c}}$ at $R\to\infty$, while larger deviations are found for small $R$.
This $R$-dependence is understandable from the structure of the momentum phase space;
the momentum here is discretized as $p_{l,k}=\xi_{l,k}/R$, but becomes continuous for large $R$ as in the usual Cartesian case.
We note that this convergence to the Cartesian limit is expected to hold for massive fields as well, since the difference arises solely from the modification of the phase-space structure.

Figure~\ref{fig:Pi_d} also shows the mode dependence of the self-energy.
The left panel exhibits that the self-energy with a larger $l$ deviates  more from the Cartesian one.
This is physically plausible because modes with large circular orbits should be more affected by the finite-size effect.
The right panel indicates that the deviation from the Cartesian self-energy is saturated for large $k$, unlike for large $l$ in the left panel.
To understand the difference, we recall the asymptotic form of the Bessel function~\cite{abramowitz1965handbook}:
\begin{equation}
 J_l(z) \approx \sqrt{\frac{2}{\pi z}} \cos{\left(z - \frac{l \pi}{2} - \frac{\pi}{4}\right)},
 \quad
 z \gg \biggl|l^2 - \frac{1}{4}\biggr|.
\end{equation}
For large $k$, hence, the zeros of the Bessel function are those of plane waves, leading to the partial equivalence between the momentum phase spaces in the cylindrical and Cartesian coordinates.
This is why $\Pi_{1\mathrm{r}}(l,k,l,k)$ for $k\gg |l|$ approaches $\Pi_{1\mathrm{c}}$ more rapidly than $k\ll|l|$. 

\begin{figure}
   \centering
    \includegraphics[width=0.5\columnwidth]{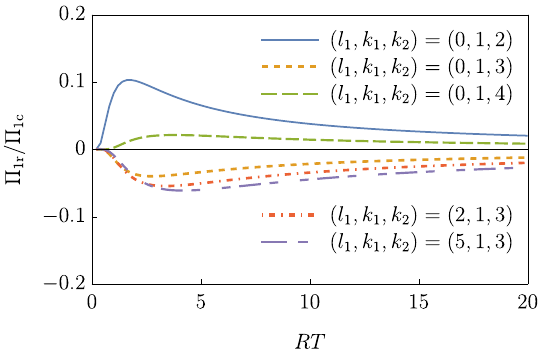}
    \caption{Off-diagonal elements of the self-energy for $\Omega = 0$.}
\label{fig:Pi_offd}
\end{figure}

In Fig.~\ref{fig:Pi_offd}, we show the off-diagonal elements of $\Pi_{1\mathrm{r}}(l_1,k_1,l_2=l_1,k_2)$ for $\Omega = m = 0$.
While the diagonal elements are always positive due to the radial integral $I_{\nu_{1\perp},\nu_{1\perp},\nu_{\perp},\nu_{\perp}} \allowbreak \sim \int \rmd r r J_{l_1}^2 J_{l}^2 > 0$, the off-diagonal one can take both signs.
The trends convergent to zero for $R\to\infty$ imply that the totally diagonal Cartesian expression~\eqref{eq:Pi_c} is reproduced.
The slower convergence for larger $l$ is, similarly to the diagonal element in Fig.~\ref{fig:Pi_d}, attributed to the larger finite-size effect for large $l$ modes.
In addition, off-diagonal elements with larger $|k_1-k_2|$ converge more quickly to zero.
This feature is consistent with the conservation of radial momentum, i.e. $p_{l,k_1}\simeq p_{l,k_2}$, at $R\to\infty$.

\begin{figure}
    \centering
    \includegraphics[width=0.5\columnwidth]{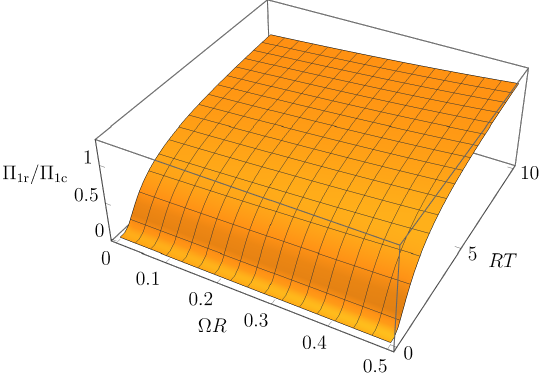}
    \caption{$\Omega$ and $T$ dependence of the diagonal self-energy.}
    \label{fig:Pi_3d}
\end{figure}

Figure~\ref{fig:Pi_3d} is the plot of the $\Omega$ dependence of the diagonal self-energy for $m=0$.
We take the lowest mode $(l,k)=(0,1)$, and the self-energy is again normalized by Eq.~\eqref{eq:Pi_c}.
In the low temperature regime $RT\ll 1$, there appears no visible rotational effect.
On the other hand, at higher temperature, the self-energy increases as a function of $\Omega$, and such an increasing behavior is enhanced more with higher temperature.
The $\Omega$ dependence at low and high temperatures above totally reflects those of the distribution function $n_\mathrm{B}(\nu)$ involved in Eq.~\eqref{eq:Pi_r}.

\section{Feynman rules and rotational effect on vacuum}
\label{sec:feyn}
From the previous computations, we read off Feynman rules in the rotating $\phi^4$ theory.
The modification compared with the usual Cartesian case is summarized as
\begin{equation}
\begin{split}
\label{eq:Frule}
\scalebox{0.8}{
 \begin{tikzpicture}[baseline={([yshift=-0.8ex]current bounding box.center)}]
 \begin{feynhand}
        \vertex (a) at (-1, 0) {\large $\nu$};
        \vertex (b) at (1, 0) {\large $\nu'$};
        \propag [plain] (a) to (b) ;
 \end{feynhand}
 \end{tikzpicture}
 }\!\!
 & =\quad \calD_0(\nu) \delta_{\nu,\nu'} ,\\
\scalebox{0.5}{
 \begin{tikzpicture}[baseline={([yshift=-0.8ex]current bounding box.center)}]
 \begin{feynhand}
        \vertex (a) at (-1, -1){\huge $\nu_1$};
        \vertex (b) at (1, 1) {\huge $\nu_3$};
        \vertex (c) at (-1, 1){\huge $\nu_2$};
        \vertex (d) at (1, -1){\huge $\nu_4$};
        \vertex [dot] (o) at (0, 0) {};
        \propag [plain] (a) to (o);
        \propag [plain] (b) to (o);
        \propag [plain] (c) to (o);
        \propag [plain] (d) to (o);
 \end{feynhand}
 \end{tikzpicture}
 } \ \
 & = 
 \begin{aligned}
 & -\lambda \beta V I_{\nu_{1\perp},\nu_{2\perp},\nu_{3\perp},\nu_{4\perp}} \\[-1.5em]
& \quad\times \delta_{\nu_{1\parallel} + \cdots + \nu_{4\parallel},0}\,\delta_{l_1+\cdots+l_4,0}.
 \end{aligned}
\end{split}
\end{equation}
The violation of translation invariance along the radial direction results in the emergence of the radial integral~\eqref{eq:I}, while those of the other quantum numbers $n$, $q$, and $l$ are respected due to the delta function.
Thanks to these Feynman rules, the diagrammatic expressions of $\ln Z$ and $\Pi$ take the same forms as that in the Cartesian coordinate, e.g.~\cite{Kapusta:2006pm}
\begin{eqnarray}
 &\ln Z_1
 = \, 3\, \scalebox{0.7}{
\begin{tikzpicture}[baseline={([yshift=-0.5em]current bounding box.center)}]
\begin{feynhand}
\vertex (a) at (-1, 0);
\vertex (b) at (1, 0);
\vertex (c) at (0, 0);
\propag [plain] (a) to [in=90, out=90,looseness=1.7] (c);
\propag [plain] (a) to [in=270, out=270,looseness=1.7] (c);
\propag [plain] (b) to [in=90, out=90,looseness=1.7] (c);
\propag [plain] (b) to [in=270, out=270,looseness=1.7] (c);
\end{feynhand}
\end{tikzpicture}
} \ ,\\
 & \Pi_1(\nu_1,\nu_2)
 =-12 \ \scalebox{0.7}{
\begin{tikzpicture}[baseline={([yshift=-0.5em]current bounding box.center)}]
\begin{feynhand}
\vertex (a) at (-1, 0) {\Large $\nu_1$};
\vertex (b) at (1, 0) {\Large $\nu_2$};
\vertex (c) at (0, 0);
\vertex (d) at (0, 1);
\propag [plain] (a) to (c);
\propag [plain] (b) to (c);
\propag [plain] (c) to [in=0, out=0,looseness=1.7] (d);
\propag [plain] (c) to [in=180, out=180,looseness=1.7] (d);
\end{feynhand}
\end{tikzpicture}
} \ ,
\end{eqnarray}
where $3$ and $12$ are from the combinatoric factors and the minus sign is because of our convention in Eq.~\eqref{eq:Frule}.

These Feynman rules are helpful in examining the property of vacuum in the perturbation theory of the rotating scalar theory.
In Eqs.~\eqref{eq:lnZ0}, \eqref{eq:lnZ-3}, and~\eqref{eq:Pi_r}, we have already observed that at zero temperature, the rotational effect is invisible up to $\mathcal{O}(\lambda)$.
This fact evokes the possibility that higher-order diagrams are also unaffected by rotation at zero temperature.
In this paper, we prove that the above conjecture is true at least up to $\mathcal{O}(\lambda^2)$;
the proof for all order is more complicated, but our argument below is applicable to higher-order corrections.

The second-order correction to the partition function is given as
\begin{equation}
\label{eq:lnZ2}
\ln Z_2 =
36\ 
\scalebox{0.7}{
\begin{tikzpicture}[baseline={([yshift=-0.5em]current bounding box.center)}]
\begin{feynhand}
\vertex (a) at (-1.5, 0);
\vertex (b) at (1.5, 0);
\vertex (c) at (-0.5, 0);
\vertex (d) at (0.5, 0);
\propag [plain] (a) to [in=90, out=90,looseness=1.7] (c);
\propag [plain] (a) to [in=270, out=270,looseness=1.7] (c);
\propag [plain] (b) to [in=90, out=90,looseness=1.7] (d);
\propag [plain] (b) to [in=270, out=270,looseness=1.7] (d);
\propag [plain] (c) to [in=90, out=90,looseness=1.7] (d);
\propag [plain] (c) to [in=270, out=270,looseness=1.7] (d);
\end{feynhand}
\end{tikzpicture}
}
+12\ 
\scalebox{0.5}{
\begin{tikzpicture}[baseline={([yshift=-0.7em]current bounding box.center)}]
\begin{feynhand}
\vertex (a) at (-1, 0);
\vertex (b) at (1, 0);
\propag [plain] (a) to [in=90, out=90,looseness=1.7] (b);
\propag [plain] (a) to [in=270, out=270,looseness=1.7] (b);
\propag [plain] (a) to [in=90, out=90] (b);
\propag [plain] (b) to [in=270, out=270] (a);
\end{feynhand}
\end{tikzpicture}
}
\ .
\end{equation}
The first diagram consists of the three-product of the same Matsubara sum~\eqref{eq:Matsubara}, and thus it is independent of $\Omega$ at zero temperature.
On the other hand, the sum in the second diagram takes a different form as
\begin{equation}
\label{eq:Matsubara_four}
 T^3\sum_{n_1+\cdots+n_4=0} \calD(\nu_1)\calD(\nu_2)\calD(\nu_3)\calD(\nu_4).
\end{equation}
Here and hereafter we impose
\begin{equation}
\label{eq:l_cons}
 l_1+l_2+l_3+l_4=0 .
\end{equation}
Converting the sums into contour integrals with $\omega_{n_i}-\rmi\Omega l_i\to -\rmi z_i$ and dropping the purely vacuum (i.e., $T=\Omega=0$) parts through renormalization, we extract the zero-temperature but finite-$\Omega$ contribution to Eq.~\eqref{eq:Matsubara_four} as
\begin{equation}
 \mathcal{I} = \Biggl[\prod_{i=1,2,3} \oint_{C_{l_i}} \frac{\rmd z_i}{2\pi\rmi} \frac{1}{z_i^2-\epsilon_{\nu_i}^2}\Biggr]\frac{1}{(z_1+z_2+z_3)^2-\epsilon_{\nu_4}^2},
\end{equation}
where $C_{l_i}$ is defined similarly to $C_l$ in Fig.~\ref{fig:contour}.
Each contour integral over $z_i$ is evaluated with their poles.
The integral $\mathcal{I}$ vanishes unless for all of $z_i$, there is at least one pole located inside of the contour $C_{l_i}$. 
Identification of pole positions requires an exhaustive case analysis of potential combinations of $l_i$ parameters.
However, due to inherent symmetry properties among the four $l_i$ variables, this analysis can be systematically reduced to three fundamental cases:
\begin{eqnarray*}
 &\mathrm{(I)}&\quad l_1,l_2,l_3\geq 0,\quad l_4\leq 0 ,\\
 &\mathrm{(II)}&\quad l_1,l_2,l_3\leq 0,\quad l_4\geq 0, \\
 &\mathrm{(III)}&\quad l_1,l_4\geq 0, \quad l_2,l_3\leq 0.
\end{eqnarray*}
In the following, we prove $\mathcal{I}=0$ in all of the three cases, implying that no rotational effect becomes visible in $\ln Z_2$ at zero temperature.

(I)(II) The integrand in the $z_1$-integral has the four poles $z_1=\pm\epsilon_{\nu_1}$ and $z_1=\zeta_1^{\pm}$ with
\begin{equation}
\label{eq:4_poles}
 \zeta_1^\pm =\pm\epsilon_{\nu_4}-z_2-z_3.
\end{equation}
It is obvious from Eq.~\eqref{eq:positivity} that the former two poles $z_1=\pm\epsilon_{\nu_1}$ are outside the contour $C_{l_1}$.
For the case (I), the positions of the latter two are identified from the following two inequalities:
\begin{equation}
\begin{split}
\label{eq:zeta1_ineq}
\mathrm{Re}\,\zeta_1^+ -\Omega l_1
&> \Omega\Bigl[|l_4|-(l_1+l_2+l_3)\Bigr] > 0 ,\\
\mathrm{Re}\,\zeta_1^-
&<-\Omega|l_4| < 0,
\end{split}
\end{equation}
where we use Eqs.~\eqref{eq:positivity}, \eqref{eq:l_cons} and
\begin{equation}
\label{eq:z_ineq}
 0\leq |\mathrm{Re}\,z_i| \leq \Omega |l_i|.
\end{equation}
Therefore, all poles in the integral over $z_1$ are outside of $C_{l_1}$, implying $\mathcal{I}=0$.
For the case (II), since the signs of $l_i$ are flipped and $C_{l_1}$ is depicted as the right contour in Fig.~\ref{fig:contour}, instead of the inequalities~\eqref{eq:zeta1_ineq} we have
\begin{equation}
\begin{split}
\mathrm{Re}\,\zeta_1^+ 
&> \Omega |l_4| > 0 ,\\
\mathrm{Re}\,\zeta_1^- -\Omega l_1
&<-\Omega\Bigl[|l_4|+(l_1+l_2+l_3)\Bigr] < 0,
\end{split}
\end{equation}
which leads to $\mathcal{I}=0$.

(III) The poles in the $z_1$-integral are the same as before, and $\mathrm{Re}\,\zeta_1^+ -\Omega l_1>0$ holds.
In this case, however, another inequality $\mathrm{Re}\,\zeta_1^-<0$ is not always satisfied, providing the finite contribution to $\mathcal{I}$ as
\begin{equation}
\begin{split}
 \mathcal{I} 
 &= \frac{-1}{2\epsilon_{\nu_4}}\Biggl[\prod_{i=2,3} \oint_{C_{l_i}} \frac{\rmd z_i}{2\pi\rmi} \frac{1}{z_i^2-\epsilon_{\nu_i}^2}\Biggr]\\
 &\quad\times \frac{\theta(-{\epsilon}_{\nu_4}-\mathrm{Re}(z_2+z_3))}{(z_2+z_3+\epsilon_{\nu_4})^2 -\epsilon_{\nu_1}^2}.
\end{split}
\end{equation}
The step function is involved so that $\mathcal{I}$ vanishes unless $\mathrm{Re}\,\zeta_1^->0$.
The integrand in the integral over $z_2$ has only three poles: $z_2=\pm\epsilon_{\nu_2}$ and $z_2=\zeta_2^-$ with
\begin{equation}
 \zeta_2^- = -\epsilon_{\nu_1} -{\epsilon}_{\nu_4}-z_3,
\end{equation}
while $z_2 = \epsilon_{\nu_1} -{\epsilon}_{\nu_4}-z_3$ is excluded by the step function.
The two poles $z_2=\pm\epsilon_{\nu_2}$ are outside of $C_{l_2}$ and the location of the last pole is found from
\begin{equation}
 \mathrm{Re}\,\zeta_2^--\Omega l_2 
 <-\Omega (|l_1| +|l_4| + l_3 + l_2) 
 = 0.
\end{equation}
Hence, the three poles in the $z_2$-integral are always outside of $C_{l_2}$, leading to $\mathcal{I}=0$.

\section{Summary}

In this paper, we developed the finite temperature perturbation theory of $\lambda\phi^4$ theory under rotation.
Due to the translation non-invariance in the radial direction, the perturbation theory is described by Bessel functions and their zeros.
The main modification in the correction to the partition function is the involvement of the integral with the four-product of Bessel functions.
The same integral also enters the one-loop self-energy, which exhibits the external-line dependence and the off-diagonal structure in contrast to the Cartesian expression.
We numerically confirmed that in the infinite-volume limit without rotation, quantities evaluated in the cylindrical coordinate correctly reproduce those in the Cartesian coordinate.

Through the Feynman rules derived in this work, we demonstrated that rotational contributions to the partition function at zero temperature completely cancel up to $\mathcal{O}(\lambda^2)$ in perturbative expansions, although extending this result to all orders remains an open problem.
This finding suggests that the perturbative vacuum is unaffected by rotation, similarly to the noninteracting theory.
In other words, we found that the \emph{vacuum sensitivity} to rotation is a nonperturbative phenomenon~\cite{Mameda:2023sst,Jiang:2023zzu,Braguta:2021jgn,Yang:2023vsw}, implying that the zero-temperature contribution is required to be carefully taken into account in the nonperturbative evaluation of thermal averages.
Unlike the perturbative vacuum, the nonperturbative vacuum structure can be significantly affected by rotation~\cite{Yamamoto:2013zwa,Braguta:2023yjn}, suggesting that the static and rotating systems may exhibit qualitatively different physical properties.
We also anticipate that the perturbative vacuum insensitivity to rotation found here holds for other interacting theories, although confirming this mathematically requires a careful analysis of the pole positions as done in this work.

In the perturbation theory of rotating scalar fields, typical modifications compared with the Cartesian one emerge only in a few Feynman rules.
In the same manner, hence, we can also compute the loop diagrams in effective models and gauge theory under rotation.
For example, applying our formulation to chiral effective models, one can evaluate quantum fluctuations and thus meson masses, which are crucial quantities in the phenomenology of chiral phase transition~\cite{Klevansky:1992qe,Hatsuda:1994pi}.
We note that the vertex of fermions and gauge field yields integrals including the three-product of Bessel functions.
Since this integral is analytically tractable~\cite{tyler1990analysis}, the perturbation theory would become more manageable.

\section*{Acknowledgements}
The authors thank Xu-Guang Huang for reading the manuscript and giving useful comments.
K.~M. is supported by JSPS KAKENHI Grant Number~24K17052, and appreciates the hospitality of Yukawa Institute for Theoretical Physics during the workshop ``Topology and Dynamics of Magneto-Vortical Matter'', where this work was discussed.

\bibliographystyle{elsarticle-num}
\bibliography{perturbation}

\end{document}